\documentclass[fleqn,usenatbib]{mnras}

\usepackage{newtxtext,newtxmath}

\usepackage[T1]{fontenc}

\DeclareRobustCommand{\VAN}[3]{#2}
\let\VANthebibliography\thebibliography
\def\thebibliography{\DeclareRobustCommand{\VAN}[3]{##3}\VANthebibliography}

\usepackage{graphicx}	
\usepackage{amsmath}	

\usepackage{comment}
\usepackage{amsmath}
\usepackage{subcaption}
\usepackage{placeins}
\usepackage{float}

\title[CH Cygni X-ray Periodicity]{A 682-second X-ray Periodicity in CH Cygni: Evidence for a Magnetic White Dwarf}

\author[Pichardo Marcano et al.]{
Manuel Pichardo Marcano,$^{1,2,3}$\thanks{E-mail:m.pichardo.marcano@astro.unam.mx}
Thomas J.  Maccarone,$^{4}$
Liliana E. Rivera Sandoval$^{5}$
\\
$^{1}$Universidad Nacional Autónoma de México. Instituto de Astronomía. A.P. 70-264, 04510. Ciudad de México, México.\\
$^{2}$Department of Life and Physical Sciences, Fisk University, 1000 17th Avenue N., Nashville, TN 37208, USA\\
$^{3}$Department of Physics \& Astronomy, Vanderbilt University, 6301 Stevenson Center Lane, Nashville, TN 37235, USA \\
$^{4}$Department of Physics and Astronomy, Texas Tech University, Lubbock, TX 79409, USA \\
$^{5}$Department of Physics and Astronomy, University of Texas Rio Grande Valley, Brownsville, TX 78520, USA\\
}

\date{Accepted XXX. Received YYY; in original form ZZZ}

\pubyear{\the\year{}}

\begin{document}
\label{firstpage}
\pagerange{\pageref{firstpage}--\pageref{lastpage}}
\maketitle

\begin{abstract}
Symbiotic stars are interacting binaries consisting of a red giant and typically a white dwarf, important as potential Type Ia supernova progenitors. Despite theoretical predictions that white dwarfs in symbiotic systems should often be magnetic, direct evidence has been elusive. We report the discovery of a $682.5 \pm 7$ s periodicity in the XMM-Newton X-ray light curve that we interpret as the spin period of the WD in CH Cygni. This detection provides strong evidence for a magnetic white dwarf in CH Cygni, making it only the second WD symbiotic star with confirmed X-ray pulsations after R Aquarii. Our discovery supports the magnetic propeller model previously proposed for CH Cygni's jet activity and state transitions. 
\end{abstract}

\begin{keywords}
white dwarf
\end{keywords}



\section{Introduction} 

Symbiotic stars (SS) are interacting binaries consisting of a cool evolved red giant star and a hot companion, typically a white dwarf (WD) accreting material from the giant's wind or via Roche lobe overflow, if they are sufficiently close. These systems are of particular importance as potential progenitors of Type Ia supernovae, either via the single degenerate scenario, if they host a massive enough WD, or via the double-degenerate channel as they are progenitors of double WD systems \citep{Kenyon1993ApJ...407L..81K,Hachisu1999ApJ...522..487H,Lu2009MNRAS.396.1086L,Liu2019A&A...622A..35L,Ilkiewicz2019MNRAS.485.5468I,Laversveiler2025A&A...698A.155L}. 

CH Cygni is among the most  studied SS, with observations spanning many wavelengths from radio \citep{Taylor1986Natur.319...38T,Crocker2002MNRAS.335.1100C} to X-rays \citep{Leahy1987A&A...176..262L,Ezuka1998ApJ...499..388E}. In all these wavelengths, CH Cygni displays complex behavior showing outbursts, variability at different timescales and state transitions in the optical \citep{Mikolajewski1990A&A...235..219M}, radio \citep{Taylor1986Natur.319...38T}, and X-ray \citep{Mukai2007PASJ...59S.177M}.

X-ray observations have proven crucial for understanding accretion processes in SS. For a review on X-rays from SS see \citet{Luna2013A&A...559A...6L} and for a review on accreting WDs in X-rays see \citep{Mukai2017PASP..129f2001M,Webb2023arXiv230310055W}. The first clear detection of CH Cygni in X-rays was made by EXOSAT in 1985 by \citet{Leahy1987A&A...176..262L}. Subsequent X-ray observations have revealed CH Cygni as a highly variable X-ray source with a complex, two-component spectrum \citep{Wheatley2006MNRAS.372.1602W}. \citet{Ezuka1998ApJ...499..388E} conducted a detailed timing analysis and detected stochastic variations on timescales as short as 100 seconds without identifying  any coherent oscillations. \citet{Mukai2007PASJ...59S.177M} reported significant hard X-ray variability using data from the Suzaku X-ray mission, and more recently, \citet{Toala2023MNRAS.522.6102T} studied high-resolution X-ray spectra to obtain chemical abundances and corroborate the presence of multitemperature X-ray-emitting gas in CH Cygni.

CH Cygni is also one of the few SS with jets seen in X-rays \citep{Brocksopp2004MNRAS.347..430B}, and its jet has been directly imaged with Chandra \citep{Galloway2004ApJ...613L..61G,Karovska2007ApJ...661.1048K,Karovska2010ApJ...710L.132K}. While the magnetic field is commonly attributed to play a key role in jet formation \citep{Stute2005A&A...432L..17S}, finding evidence of magnetic WDs in SS has been elusive. The difficulty arises partly because the red giant's contribution can mask or dilute signals from the accreting WD. Optical observations, which have been the primary tool for searching for magnetic WDs in SS, require both high cadence and millimagnitude precision to detect the periodic modulations expected from rotating magnetic WDs.

For CH Cygni, using optical ground-based data, \citet{Mikolajewski1990AcA....40..129M} reported the detection of an oscillation with a period of $\sim 500$ s, which they attributed to the rotation of a magnetized ($\sim 10$ MG) WD, proposing a magnetic propeller model to explain the system's jets, luminosity changes and flickering \citep{Mikolajewski1988ASSL..145..233M,Mikolajewski1990AcA....40..129M,Panferov2000astro.ph..7009P}. However, subsequent observations failed to confirm this periodicity. Neither \citet{Hoard1993PASP..105.1232H} nor \citet{Rodgers1997PASP..109.1093R} detected the $\sim 500$ s signal, leading to questioning on the  magnetic interpretation for CH Cygni \citep[e.g.][]{Sokoloski2003ApJ...584.1027S}.

Optical periodicities attributed to magnetic WDs have been confirmed in other SS. The first persistent detection was a $1682.6 \pm 0.6$ s periodicity in Z Andromedae \citep{Sokoloski1999ApJ...517..919S}, later confirmed by \citet{Gromadzki2006AcA....56...97G}. Since then, only a handful of additional magnetic WD candidates have been identified through optical observations: BF Cyg \citep{Formiggini2009MNRAS.396.1507F}, Hen 2-357 \citep{Toma2016MNRAS.463.1099T}, FN Sgr \citep{Magdolen2023A&A...675A.140M}, and most recently AE Cir and CI Cyg discovered using TESS \citep{Merc2024A&A...683A..84M,Merc2025arXiv250416825M}. Currently, only five symbiotic systems show periodic modulations shorter than one hour attributed to magnetic WD rotation. Notably, all except CH Cygni exhibit ellipsoidal variations, indicating their red giants nearly fill their Roche lobes. 

While optical searches have yielded these detections, X-ray observations provide more direct evidence for magnetic accretion. When the WD magnetic field exceeds $10^{5-6}$ G at the surface, X-rays arise from magnetically channeled accretion onto localized regions of the WD, producing modulation at the spin period. Despite this diagnostic power, coherent X-ray oscillations from SS have proven extremely elusive. Only one prior detection exists: \citet{Nichols2007ApJ...660..651N} discovered a 1734 s modulation in R Aquarii's (R Aqr) hard X-ray flux from 2004 Chandra observations, attributing it to a magnetic WD.

In this paper, we report the discovery of a $682.5 \pm 7$ s modulation in X-ray observations of CH Cygni obtained with XMM-Newton in 2018. This represents only the second detection of coherent X-ray pulsations from a WD SS and the first for CH Cygni. The  detection provides compelling evidence that CH Cygni indeed harbors a magnetic WD, validating aspects of the propeller model proposed three decades ago. 

\section{Data Analysis and Results}
\subsection{XMM-data}

CH Cygni was observed by XMM-Newton on 2018 May 24 (Obs. ID: 0830190801; PI: N. Schartel). The observations were obtained using the thick optical blocking filter with the EPIC cameras in small window mode (0.15 to 15 keV). The total exposure time for the EPIC-pn, MOS1, and MOS2 cameras was 34.3, 34.7, and 34.7 ks, respectively. The average count rates, calculated using bin times of 0.12 s for EPIC-pn and 10 s for MOS1 and MOS2, were $3.76\pm 0.013$, $0.96 \pm 0.0054$, and $0.99 \pm 0.0055$ counts s$^{-1}$, respectively.

For our timing analysis, we retrieved the pipeline-processed source light curves from the XMM-Newton Science Archive for the EPIC camera.

\subsection{Timing Analysis}

\begin{figure*}
\centering
\begin{subfigure}[b]{0.48\textwidth}
    \includegraphics[width=\textwidth]{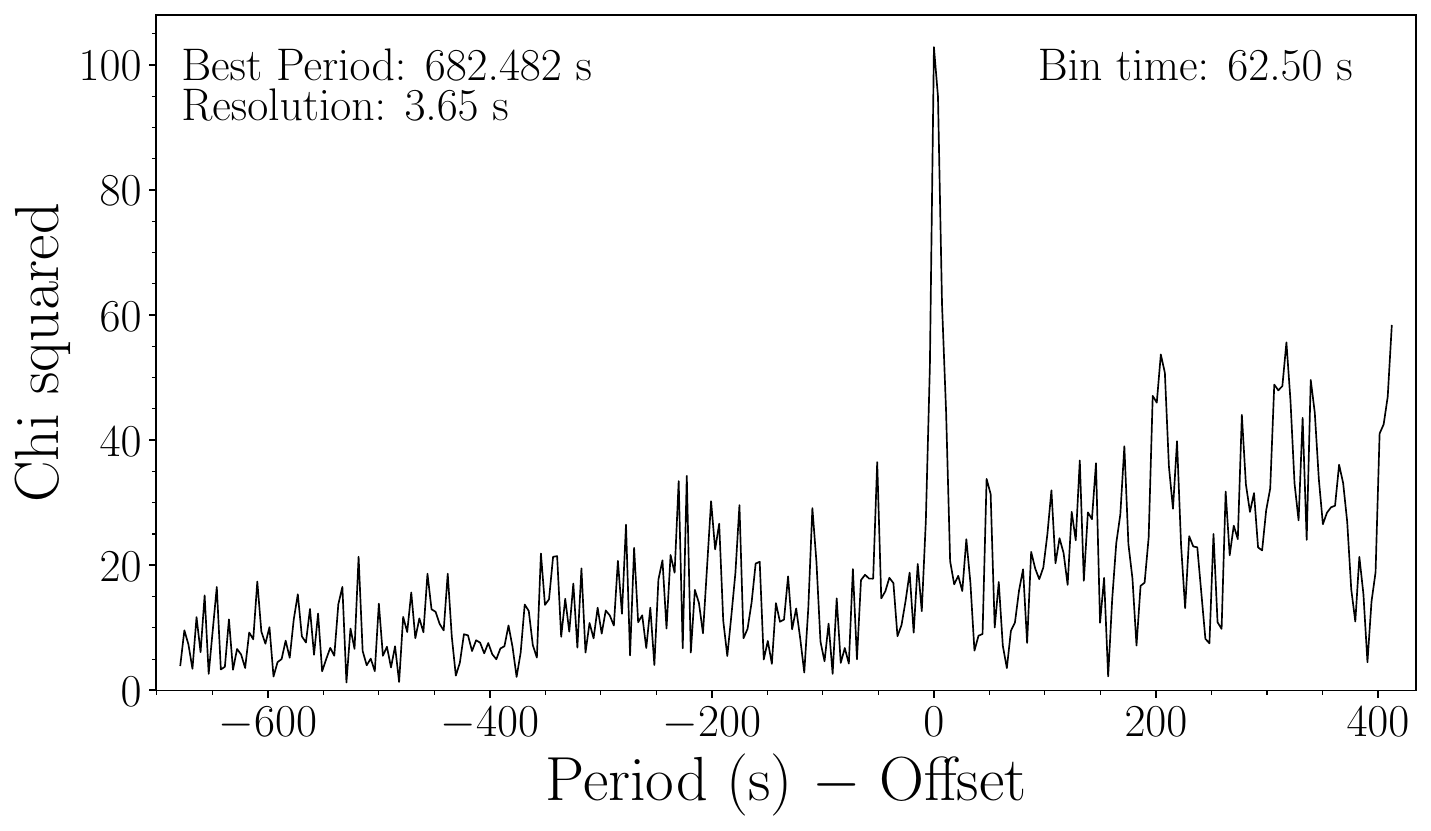}
    \label{fig:efsearch}
\end{subfigure}
\hfill
\begin{subfigure}[b]{0.48\textwidth}
    \includegraphics[width=\textwidth]{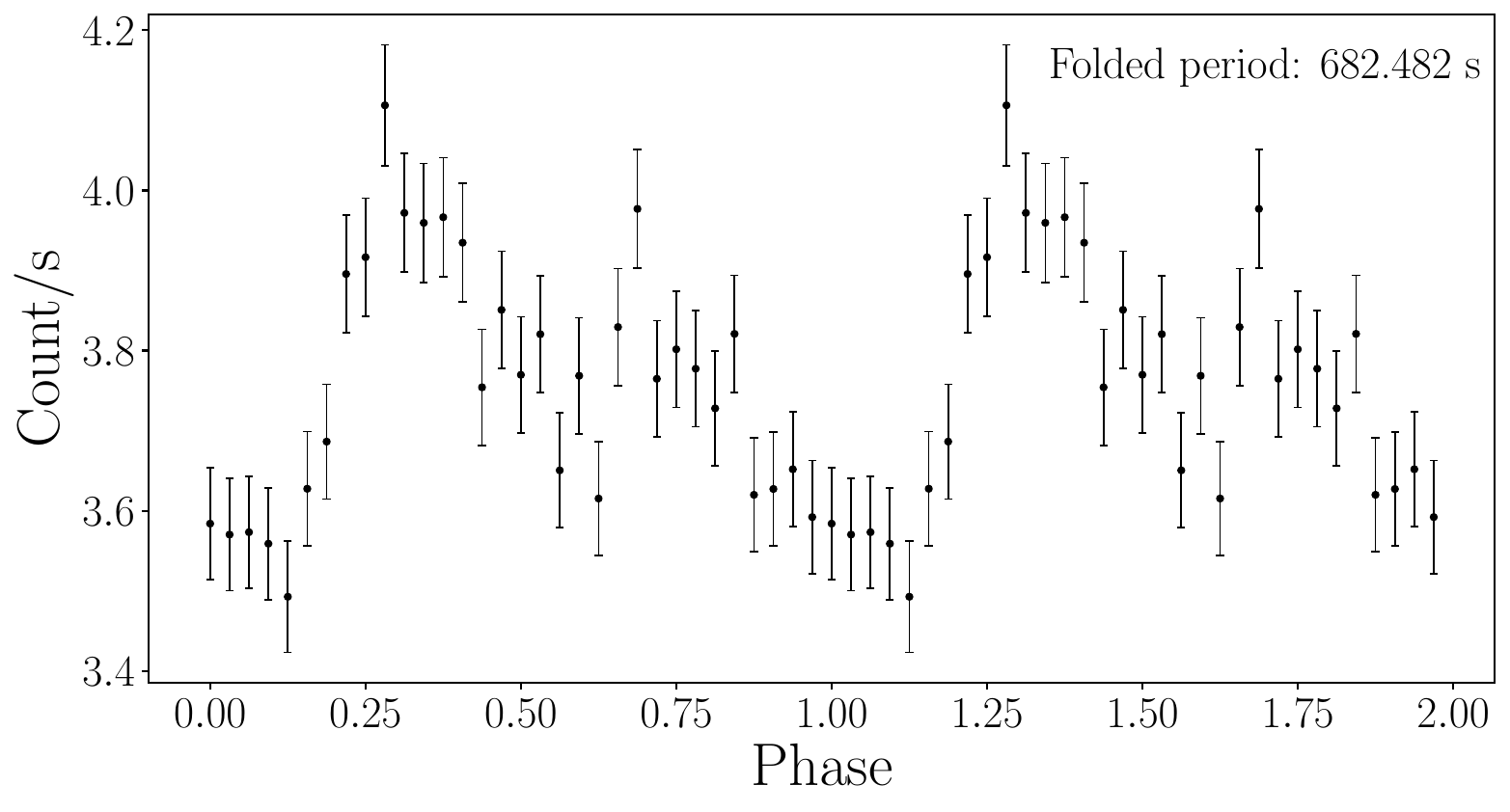}
    \label{fig:efold}
\end{subfigure}
\caption{Left: $\chi^2$ statistic as a function of trial period from the epoch folding search of the XMM-Newton EPIC-pn light curve. A highly significant isolated peak is detected at 682.5 s. Right: XMM-Newton EPIC-pn light curve of CH Cygni folded at the 682.5 s period using 32 phase bins. Two cycles are shown for clarity. A marginal secondary peak is visible near phase 0.7.}
\label{fig:period_analysis}
\end{figure*}

We searched for periodic signals using the epoch folding technique implemented in the XRONOS\footnote{\url{https://heasarc.gsfc.nasa.gov/xanadu/xronos/xronos.html}} package \citep{Leahy1983ApJ...266..160L}, which folds the light curve at trial periods and computes the $\chi^2$ statistic for the deviation of the folded profile from a constant. We found a significant peak at $682.5 \pm 7$ s (Figure~\ref{fig:period_analysis}) with $\chi^2 = 103$. For $N-1 = 7$ degrees of freedom, where $N = 8$ is the number of phase bins, and 300 trial periods, we get a false alarm probability of $10^{-17}$. This represents the probability that the profile has been produced by noise, and the likelihood that the X-ray count rate is not constant.  The uncertainty on the period  was estimated from the width of a Gaussian fit to the $\chi^2$ distribution around the peak. The folded light curve at this period (Figure~\ref{fig:period_analysis}) displays a clear modulation with a marginal secondary peak at phase $\sim 0.7$. The  Fourier power spectrum using the XRONOS tool POWSPEC is shown in the Appendix in Figure~\ref{fig:pw}.

We performed the epoch folding period search exclusively on the EPIC-pn light curve due to its higher count rate compared to the MOS cameras. For consistency check, we  folded the MOS1 and MOS2 light curves at the detected 682.5 s period and confirmed the presence of the periodicity. All subsequent analysis focuses on the EPIC-pn data.


\section{Discussion}

\subsection{The magnetic field of CH Cygni}

The detection of a 682.5 s periodicity in the X-ray emission provides strong support for magnetic accretion in CH Cyg. We can constrain the WD's magnetic field strength using two  approaches.

For magnetic accretion to occur, the WD's magnetic field must be strong enough to channel the accretion flow onto the WD surface. This requires the magnetospheric radius, $r_{\mathrm{mag}}$, to exceed the WD radius, $R$, where $r_{\mathrm{mag}}$ is defined as the distance at which the magnetic pressure of the WD's dipolar field balances the ram pressure of the accreting matter. This condition yields a minimum field strength:

\begin{multline}
B_{min} \gtrsim  10^4 G \left (  \frac{R}{10^{9}\; \mathrm{cm}} \right )^{-5/4} \left (  \frac{\dot M}{10^{-8} \;\mathrm{M_{\odot} \;yr^{-1}}} \right )^{1/2} \\
    \times \left (  \frac{M_{\mathrm{wd}}}{0.6 \; M_{\mathrm{wd}}} \right )^{1/4}
\end{multline}

Another constraint comes from assuming that the WD has reached spin equilibrium, where $r_{\mathrm{mag}}$ equals the corotation radius, $R_{\Omega}$. The corotation radius for the observed 682 s period is:

\begin{equation}
R_{\Omega} = \left ( \frac{G M P_{\mathrm{spin}}^2}{4 \pi^2} \right )^{1/3} = 9.8 \times 10^{9} \; \mathrm{cm}
\end{equation}

Setting $r_{\mathrm{mag}} = R_{\Omega}$, we derive an equilibrium magnetic field strength:

\begin{multline}
B_{eq} \approx 10^6 \; G   \left (  \frac{M_{\mathrm{wd}}}{0.6 \; M_{\mathrm{wd}}} \right )^{5/6} 
\left (  \frac{\dot M}{10^{-8} \;\mathrm{M_{\odot} \;yr^{-1}}} \right )^{1/2} \\
\times \left (  \frac{R}{10^{9}\; \mathrm{cm}} \right )^{-3} 
 \left (  \frac{P_{\mathrm{spin}}}{682\; \mathrm{s}} \right )^{7/6} 
\end{multline}

To verify that spin equilibrium is achievable, we calculate the timescale required to spin up the WD from rest to the observed 682 s period. The accretion torque is:

\begin{equation}
N = \dot M (G \times M_{wd}  r_{mag}) ^{1/2}
\end{equation}

For a WD with moment of inertia $I \sim 10^{50} \mathrm{g}\, \mathrm{cm^2}$, the spin-up timescale becomes:

\begin{equation}
 \mathrm{t_{spin-up}} = \frac{2 \pi I }{5 N P_{spin}} = 2 \times 10^4 \mathrm{yr} 
\end{equation}

This timescale is shorter than the red giant lifetime ($\sim 10^6$ yr), confirming that the WD can reach spin equilibrium during the current mass transfer phase.



\begin{table*}
\centering
\caption{X-ray observations of symbiotic stars: search for periodicities.}
\label{tab:xray_nondetections}
\begin{tabular}{lcccccccl}
\hline
Object & Date (Instrument) & $F_{\rm X}$ & $L_{\rm X}$ & Count Rate & Exp. & RMS & SNR$_{\rm var}$ & Comments \\
 & & (10$^{-11}$ cgs) & (10$^{32}$ erg s$^{-1}$) & (ct s$^{-1}$) & (ks) & & & \\
\hline
\multicolumn{9}{c}{\textit{Periodicity Detections}} \\
\hline
CH Cyg & 2018 May (XMM) & 7.5 [0.2--10.0 keV] & 3.5 & 3.8 & 34.3 & 0.04 & 102 & 682.5 s periodicity [This work] \\
R Aqr & 2004 Jan (Chandra) & 0.02 [0.3--8 keV]$^a$ & 0.01 & 0.003 & 40 & 0.5 & 15 & 1734 s periodicity$^a$ \\
\hline
\multicolumn{9}{c}{\textit{No Periodicity Detected}} \\
\hline
CH Cyg & 1994 Oct (ASCA) & 6.9 [2--10 keV]$^b$ & 3.3 & 0.51 & 25 & $<0.03^\dagger$ & (5) & No coherent periodicity detected$^b$ \\
RT Cru & 2005 Oct (Chandra) & 0.91 [0.3--8 keV]$^c$ & 43 & 0.11 & 52 & $<0.04^\dagger$ & (5) & Hard X-ray source \\
RT Cru & 2023 Feb (XMM) & 1.39 [0.1--15 keV]$^d$ & 66 & 0.8 & 59 & $<0.01^\dagger$ & (5) & Hard X-ray source \\
MWC 560 & 2007 Sept (XMM) & 0.02 [0.1--15 keV]$^d$ & 1.4 & 0.008 & 34.4 & $<0.2^\dagger$ & (5) & Hard X-ray source with a jet \\
\hline
\end{tabular}
\begin{flushleft}
\footnotesize
\textbf{Notes}---SNR$_{\rm var} = \frac{1}{2}Ir^2(\frac{T}{\lambda})^{0.5}$ where I is the intensity, in counts per unit time, r is the fractional root-mean-squared (rms) amplitude of the variability, T is the exposure time and $\lambda$ is the frequency width over which the variation is measured. We assume $\lambda = 1/T$. Luminosities calculated using distances: CH Cyg (200 pc), RT Cru (2 kpc), MWC 560 (2.5 kpc), R Aqr (200 pc). $^\dagger$ Upper limit assuming a detection limit of SNR$_{\rm var} = 5$. 
$^a$ J. S. Nichols et al. (2007). $^b$ H. Ezuka et al. (1998). $^c$ G. J. M. Luna \& J. L. Sokoloski (2007). $^d$ XMM Archive.
\end{flushleft}
\end{table*}

\subsection{Previous periodicity  searches}

Our detection of X-ray pulsations from CH Cygni provides compelling evidence for coherent periodicity in this system. Previous searches have produced conflicting results across different wavelengths and epochs, with most failing to confirm persistent periodic behavior.

The first claim of periodic behavior came from \citet{Mikolajewski1990AcA....40..129M}, who reported optical pulsations with a period near 500 s and proposed that CH Cygni harbors a magnetic WD. However, subsequent optical studies failed to confirm this detection. \citet{Hoard1993PASP..105.1232H} and \citet{Rodgers1997PASP..109.1093R} found no evidence of periodicity between 500 and 600 s in their optical photometry, though they reported possible periods at approximately 2200 and 3000 s. \citet{Sokoloski2003ApJ...584.1027S} came to the same conclusion and did not find any evidence for periodic or quasi-periodic oscillations in the optical emission from CH Cygni in either the high or low state, casting doubt on the magnetic propeller model. Prior X-ray searches similarly yielded null results. \citet{Ezuka1998ApJ...499..388E} analyzed ASCA data and found no significant coherent power between 32 and 65536 s, despite the earlier optical claims. This absence of X-ray pulsations, a key signature of magnetic accretion onto WDs, had cast doubt on whether magnetic processes dominate the accretion in CH Cyg.

The discovery of an X-ray periodicity of 682.5 s provides strong evidence for magnetic accretion in CH Cyg. This detection  makes CH Cygni only the second WD SS with detected X-ray pulsations potentially from a WD, after R Aqr \citep{Nichols2007ApJ...660..651N}. The folded light curve (Figure~\ref{fig:period_analysis}) shows a double-peaked pulse profile, which may suggest an accretion geometry involving two magnetic poles. However, the second peak is very marginal and should be interpreted with caution. A similar, though more prominent, double-humped profile has been observed in another magnetic WD candidate, BF Cyg \citep{Formiggini2009MNRAS.396.1507F}.

Since earlier X-ray observations failed to detect this periodicity, future X-ray monitoring campaigns are crucial to determine whether the 682.5 s oscillation persists across different epochs. Multi-epoch observations with XMM-Newton, Chandra, or future X-ray missions would allow us to track potential period derivatives, search for correlations between pulse amplitude and accretion rate, and investigate whether the marginal double-peaked profile remains stable. Such observations could also reveal whether the periodicity disappears during quiescent states, which would provide strong support for the magnetic propeller model's predictions about state-dependent accretion geometry.


\subsection{CH Cygni as a Magnetic Propeller}

\citet{Mikolajewski1988ASSL..145..233M,Mikolajewski1990A&A...235..219M} invoked a magnetic propeller model based on the oblique rotator theory of \citet{Lipunov1987Ap&SS.132....1L}. This has been explored further by \citet{Panferov2000astro.ph..7009P}, where they suggest the presence of a strongly magnetized WD ($B\sim 10^6$ G) to account for the jets, rapid flickering, and large luminosity variations observed in CH Cyg. 

Support for the magnetic propeller interpretation also comes from time-resolved spectroscopy. \citet{Tomov1996MNRAS.278..542T} identified evidence for a magnetic propeller state in 1994, characterizing three distinct accretion phases based on the accretion rate: quiet (inactive), low, and high. In the high or accretor phase, material overcomes the magnetic barrier and establishes polar-column accretion onto the WD surface.

CH Cygni exhibits dramatic X-ray variability, with flux changes exceeding an order of magnitude between observations \citep{Mukai2007PASJ...59S.177M,Mukai2009ATel.2046....1M,Mukai2009ATel.2245....1M}. During the 2018 XMM observation when we detected the periodicity, CH Cygni was in a high state with a total X-ray flux of $7.5 \times 10^{-11}$ erg s$^{-1}$ cm$^{-2}$ \citep[0.2–10.0 keV;][]{Toala2023MNRAS.522.6102T}. This flux is comparable to that observed by ASCA over 20 years earlier, which reported $6.9 \times 10^{-11}$ erg s$^{-1}$ cm$^{-2}$ \citep[2–10 keV; ][]{Ezuka1998ApJ...499..388E}. Intriguingly, despite the comparable flux levels, \citet{Ezuka1998ApJ...499..388E} reported no evidence for coherent periodicity in the ASCA data. We applied the same epoch folding technique to the ASCA data and found no periodicity. We folded the ASCA X-ray light curve on the 682.5 s period and present the result in the Appendix (Figure~\ref{fig:ascafolded}). The ratio of measured to expected variance in the binned light curve is 2.19, and the folded profile shows $\sim 1$–$2\%$ variability in the count rate, similar to that seen in the XMM-Newton data (Figure~\ref{fig:period_analysis}). It is therefore possible that the periodicity was present in the ASCA observations but detectable only with XMM-Newton owing to its larger effective area. Future monitoring will be essential to determine whether the detectability of the oscillation depends on more than the accretion rate alone and to test whether the presence of X-ray pulsations correlates with specific accretion states in CH Cygni, providing critical constraints on the magnetic propeller model.







\subsection{Comparison with Other SS}

CH Cygni is not unique in displaying jet activity and claims of periodic variability. MWC 560 represents a similar system to CH Cygni and has also being proposed as a system holding a magnetic propeller \citep{Panferov2000astro.ph..7009P}. It also exhibits prominent jet activity and optical quasi-periodic oscillations, including semicoherent 22-minute variations \citep{Dobrzycka1996AJ....111..414D,Gromadzki2006AcA....56...97G}. Like CH Cygni, periodicities have been claimed and used as evidence for a magnetic propeller model \citep{Michalitsianos1993ApJ...409L..53M}. However, X-ray observations of MWC 560 ($L_X = 1.4 \times 10^{32}$ erg/s [0.1-15 keV]) revealed only aperiodic variability on timescales of minutes to hours, with no periodic rapid variability detected \citep{Stute2009A&A...498..209S}. The absence of X-ray pulsations in MWC 560, despite its other similarities to CH Cygni, highlights the rarity of detecting coherent X-ray modulations in SS.


R Aqr shares remarkable similarities with CH Cygni, exhibiting both X-ray periodicity and jet activity detected at X-ray wavelengths. \citet{Nichols2007ApJ...660..651N} discovered a 1734 s modulation in hard X-ray flux, making it the first symbiotic with confirmed X-ray pulsations potentially from a magnetic WD. The presence of both X-ray pulsations and jets in R Aqr, now also seen in CH Cygni, suggests a connection between magnetic WDs and jet formation in symbiotic systems.

Z Andromedae, while showing optical periodicity at 1682.6 s \citep{Sokoloski1999ApJ...517..919S} and radio jet activity \citep{Brocksopp2004MNRAS.347..430B}, has not yet shown X-ray pulsations. This system demonstrates that optical periodicities do not necessarily guarantee detectable X-ray modulations.

CH Cygni belongs to a small class of symbiotic stars that produce hard X-rays \citep[$>20$ keV;][]{Kennea2009ApJ...701.1992K}. This exclusive group includes RT Cru \citep{Chernyakova2005ATel..519....1C,Bird2007ApJS..170..175B,Luna2007ApJ...671..741L}, T CrB \citep{Tueller2005ATel..669....1T}, and CD-57 3057 \citep{Masetti2006ATel..715....1M,Smith2008PASJ...60S..43S}. This may indicate that these systems host comparably massive WDs, since the maximum plasma temperature is set by the depth of the WD's gravitational potential well \citep{Byckling2010MNRAS.408.2298B}. In other accreting WD binaries like cataclysmic variables, hard X-ray emission is typically associated with magnetic WDs, where the X-rays are modulated at the WD spin period \citep{Shaw2020MNRAS.498.3457S}. However, this correlation does not appear to hold for SS. \citet{Luna2007ApJ...671..741L} searched  for X-ray pulsations from RT Cru without success, concluding that its magnetic field, if present, is insufficient to disrupt the accretion flow ($B < 10^4$ G). Our detection in CH Cyg, combined with the non-detection in RT Cru, suggests that hard X-ray emission in SS may arise through multiple mechanisms, not all requiring strong magnetic fields capable of channeling accretion.



To assess the detectability of periodic X-ray signals in SS similar to CH Cyg, we calculated the signal-to-noise of the variations, $SNR_{\mathrm{var}}$, following \citet{vanderKlis1989} \citep[see also][]{Maccarone2019RNAAS...3..116M}. SNR$_{\mathrm{var}} = \frac{1}{2}Ir^2(T/\lambda)^{0.5}$, where $I$ is the count rate, $r$ is the fractional root-mean-squared (RMS) amplitude, $T$ is the exposure time, and $\lambda$ is the the frequency width over which the variation is measured, we assume $\lambda= 1/T$. Adopting a detection threshold of SNR$_{\mathrm{var}} = 5$, we derive RMS upper limits for non-detections of periodicities. For detections we calculated the RMS by integrating the fractional rms-squared normalized power spectrum using Stingray \citep{Huppenkothen2019ApJ...881...39H,bachettiStingrayFastModern2024} \citep[see also][]{Belloni1990A&A...227L..33B,Miyamoto1992ApJ...391L..21M}. We list these values in Table~\ref{tab:xray_nondetections}.

CH Cyg exhibited 4\% RMS variability in 2018 (SNR$_{\mathrm{var}}$ = 102) but showed no periodicity in the 1994 ASCA observation \citep{Ezuka1998ApJ...499..388E}, which constrains variability to $<$3\% RMS. Given XMM-Newton's larger effective area, the periodicity may have been present but below ASCA's detection threshold (see Fig.~\ref{fig:ascafolded}). Alternatively, the lack of detection in ASCA could indicate that X-ray periodicity in SS is a transient phenomenon, dependent on specific accretion states or geometry where the magnetospheric radius permits channeled flow onto magnetic poles. The variability upper limits for other systems support this interpretation. RT Cru observations in 2023 yielded upper limits of 4\% and 1\% RMS respectively, while MWC 560's low count rate (0.008 counts per second) requires 20\% RMS variability to reach our detection threshold. In contrast, R Aqr showed 50\% RMS variability with Chandra \citep[SNR$_{\mathrm{var}}$ = 15;][]{Nichols2007ApJ...660..651N}.

The diversity of periodic and jet properties among these systems suggests that while magnetic WDs may be common in jet-producing SS, the detectability of their signatures depends on multiple factors including accretion rate, magnetic field strength, and system orientation. CH Cygni's newly discovered X-ray periodicity places it in an exclusive group with R Aqr as the only SS showing both X-ray pulsations and jets, strengthening the connection between magnetic accretion and collimated outflows in these systems.

\section{Conclusion}

We report the discovery of a $682.5 \pm 7$ s periodicity in the X-ray emission of CH Cygni from 2018 XMM-Newton observations. We interpret this as the spin period of a magnetic WD, making CH Cygni only the second symbiotic star with detected X-ray pulsations after R Aqr.

This X-ray periodicity provides strong support for the magnetic propeller model proposed close to three decades ago by \citet{Mikolajewski1988ASSL..145..233M,Mikolajewski1990AcA....40..129M}. 

The identification of magnetic WDs in SS has broader implications for understanding WD binaries and the origin of WD magnetic fields \citep{Belloni2024A&A...686A.226B,Maccarone2024MNRAS.529L..28M}.  Future X-ray monitoring will be essential to track any period evolution and establish whether the pulsations persist across different accretion states.

\section*{Acknowledgements}
We are grateful to the anonymous referee for the careful reading and helpful suggestions. MPM acknowledges support from the EMIT NSF grant (NSF NRT-2125764). MPM is partially supported by the Swiss National Science Foundation IZSTZ0\_216537 and by UNAM PAPIIT-IG101224. This research has made use of the Science Explorer (SciX), funded by NASA under Cooperative Agreement 80NSSC21M00561. MPM would like to thank Kelly Holley-Bockelmann, his family Clemencia Marcano, Matilde Marcano, and Fritz Pichardo, his officemate Andrea Derdzinski, and the whole EMIT and Vanderbilt astronomy community for their support.

This research made use of Astropy \citep{2013A&A...558A..33A,2018AJ....156..123A,2022ApJ...935..167A}, HEAsoft \citep{Heasarc2014ascl.soft08004N}, Stingray \citep{Huppenkothen2019ApJ...881...39H,bachettiStingrayFastModern2024}.

\section*{Data Availability}

This work used public data available in the XMM-Newton Science Archive (XSA) under the programme ID:0830190801 (PI: N. Schartel).



\bibliographystyle{mnras}
\bibliography{sample701} 




\appendix

\FloatBarrier

\section{XMM-Newton Power Spectrum and ASCA Light Curve}

\begin{figure}
\centering
\begin{minipage}{0.48\textwidth}
    \centering
    \includegraphics[width=\textwidth]{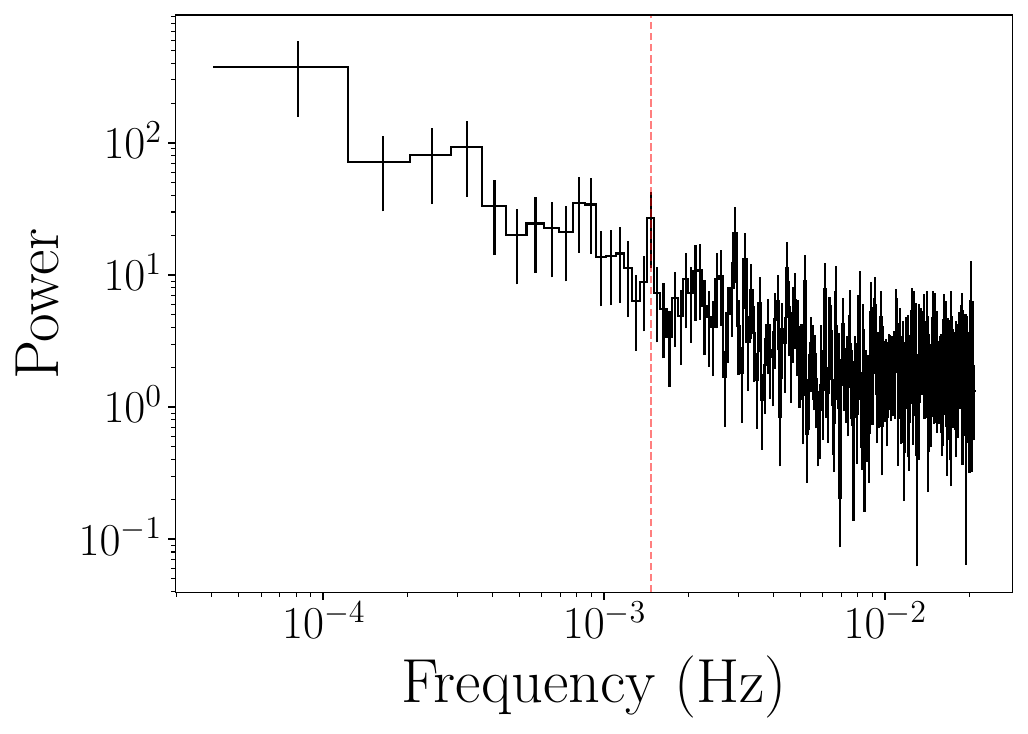}
    \caption{Power spectrum of the XMM-Newton EPIC light curve of CH Cygni. The spectrum is shown in Leahy normalization. The red line marks the detected periodicity at 682.5 s.}
    \label{fig:pw}
\end{minipage}
\hfill
\begin{minipage}{0.48\textwidth}
    \centering
    \includegraphics[width=\textwidth]{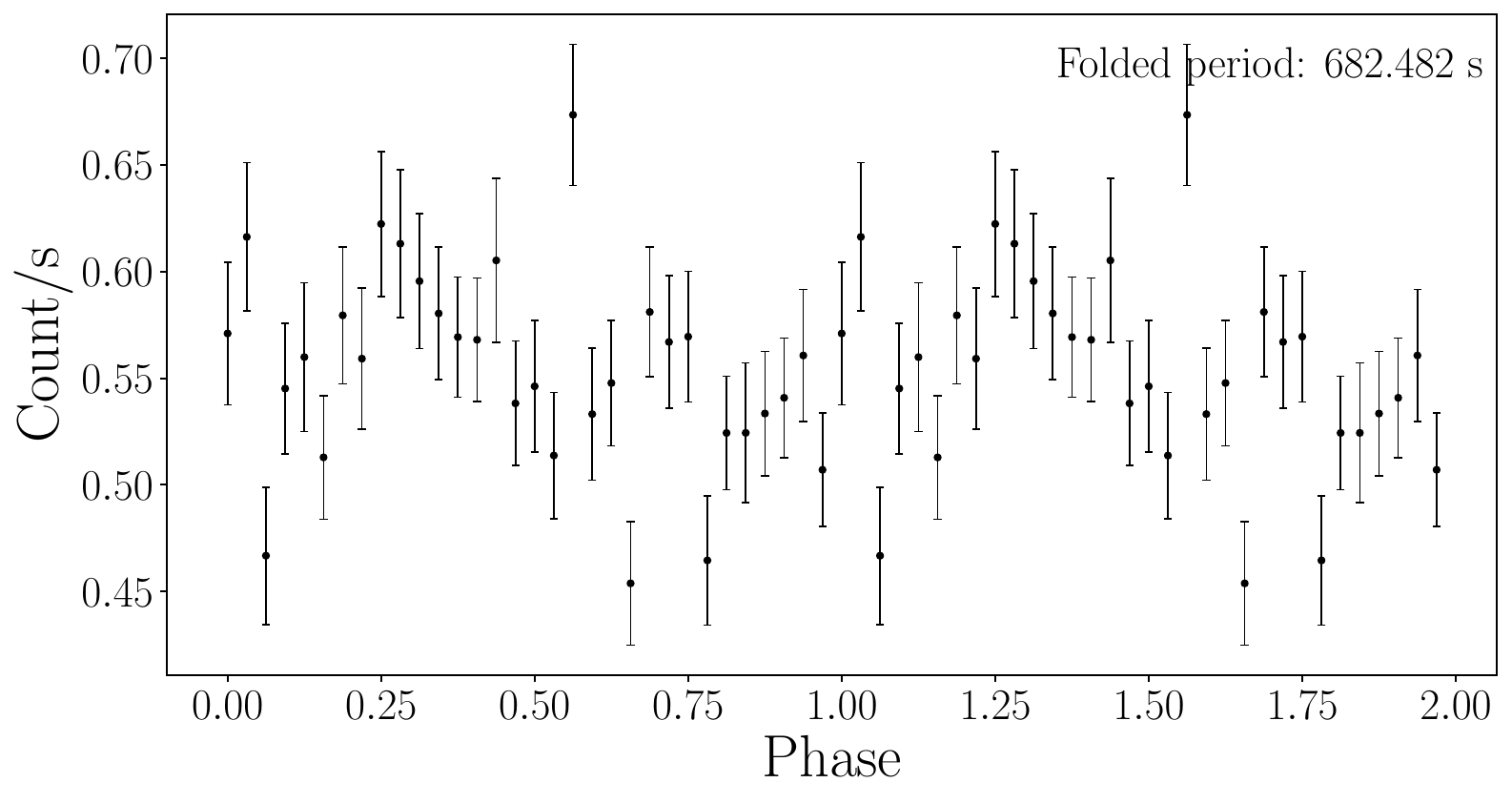}
    \caption{ASCA X-ray light curve of CH Cygni folded on the 682.5 s period detected with XMM-Newton using 32 phase bins. The profile shows marginal variability at the $1–2\%$ level, similar to that seen in the XMM-Newton data.}
    \label{fig:ascafolded}
\end{minipage}
\end{figure}

We computed the power spectrum of the XMM-Newton EPIC light curve using the XRONOS tool POWSPEC with Leahy normalization. The resulting power spectral density is shown in Figure~\ref{fig:pw}. A peak is present at 682.5 s (red dashed line), corresponding to the periodicity detected in the epoch-folding analysis.


We folded the ASCA X-ray light curve on the 682.5 s period derived from the XMM-Newton data. The folded profile, shown in Figure~\ref{fig:ascafolded}.  

\bsp	
\label{lastpage}
\end{document}